\documentclass[a4paper,11pt]{article}
\usepackage{pos}
\usepackage{xspace}
\usepackage{floatflt}

\newcommand*{\sqs}{\ensuremath{\sqrt{s}}\xspace}

\newcommand*{\kT}{\ensuremath{k_\mathrm{T}}\xspace}

\newcommand*{\Lcp}{\ensuremath{\Lambda_\mathrm{c}^+}\xspace}

\title{Jet measurements with ALICE: \\ substructure, dead cone, charm jets}
\ShortTitle{Jet measurements with ALICE}

\author*[a,1]{Róbert Vértesi}

\affiliation[a]{Wigner Research Centre for Physics, MTA Centre of Excellence, \\
  29-33 Konkoly-Thege Miklós út, 1121 Budapest, Hungary}

\note{For the ALICE collaboration.}

\emailAdd{robert.vertesi@cern.ch}

\abstract{A selection of recent jet measurements are presented from the ALICE experiment at the CERN LHC in proton-proton collisions at $\sqrt{s}=13$ TeV, focusing on substructure results for inclusive and charmed jets. The groomed jet momentum fractions ($z_g$) of  inclusive full jets are shown for various jet resolution parameters, and $z_g$, the groomed splitting radius ($\theta_g$) as well as the number of soft drops ($n_{\rm SD}$) of inclusive and charmed charged-particle jets are compared. The first direct measurement of the dead cone in heavy-flavor jets is also presented. Furthermore, the parallel momentum fractions of charmed D$^0$ mesons and \Lcp baryons are shown. Besides serving as a reference for jet structure modification measurements in heavy-ion collisions, these results provide new insight to QCD parton shower properties and flavor-dependent fragmentation processes.}

\FullConference{%
  The Eighth Annual Conference on Large Hadron Collider Physics-LHCP2020 \\
  25-30 May, 2020\\
  online}

\begin{document}
\maketitle

\section{Introduction}

Measurement of jets in small collision systems serve as fundamental tests of pQCD and hadronization models~\cite{Kang:2019prh}. Furthermore, these measurements provide a baseline for modification of jet production rates and structures in heavy-ion collisions by their interactions with the medium that is present in such collisions~\cite{Ringer:2019rfk}. Identification of jets with heavy flavor allows for the investigation of flavor-dependent production stemming from mass and color-charge effects, and the understanding of mass-dependent fragmentation.
In this contribution, recent groomed substructure measurements of inclusive and heavy-flavor jets are presented, as well as the first direct measurement of the dead cone, and the parallel momentum fraction of charmed D$^0$ mesons and \Lcp baryons. Approximately 59 pb$^{-1}$ integrated luminosity is used from pp collisions at $\sqrt{s}=13$ TeV. Charged-particle jets are reconstructed in the ALICE~\cite{Abelev:2014ffa} central barrel in the central pseudorapidity region $|\eta|<0.9$ from tracks identified in the Time Projection Chamber and the Inner Tracking System (ITS). Full jets are reconstructed in a more limited acceptance within the pseudorapidity and azimuthal angle range $|\eta|<0.7$ and $1.4<\varphi<\pi$, also using information from the Electromagnetic Calorimeter. Heavy-flavor hadrons are fully reconstructed from their decays (in the D$^0/\overline{{\rm D}^0}\rightarrow$K$^\pm\pi^\mp$ and $\Lcp\rightarrow$pK$^0_S$ channels), aided by ITS based on statistical selection of tracks originating from a secondary vertex.

\section{Groomed substructure of inclusive jets}

\begin{figure}
	\centering
	\includegraphics[width=0.4\textwidth]{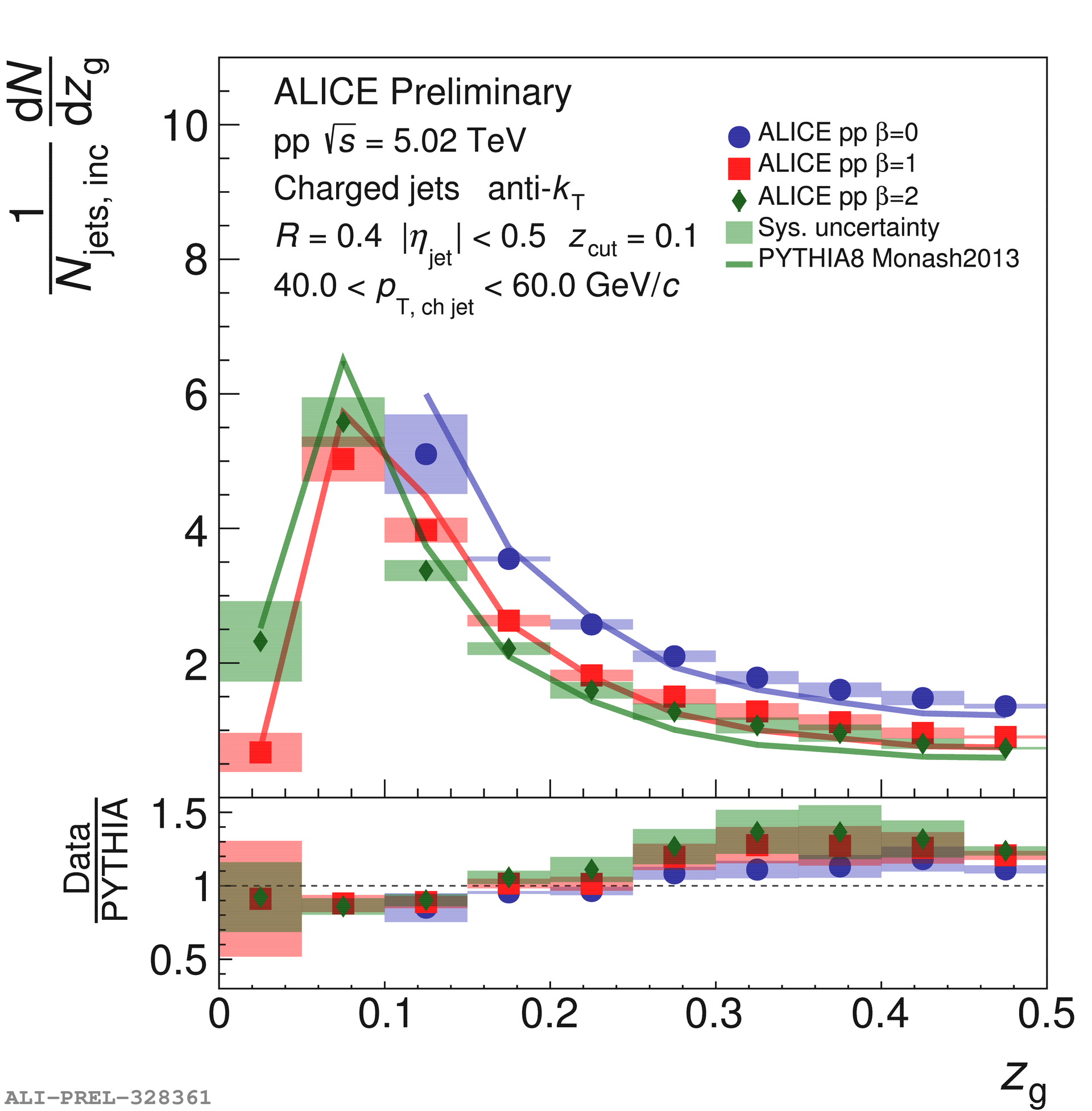}
	\hspace{0.05\textwidth}
	\includegraphics[width=0.4\textwidth]{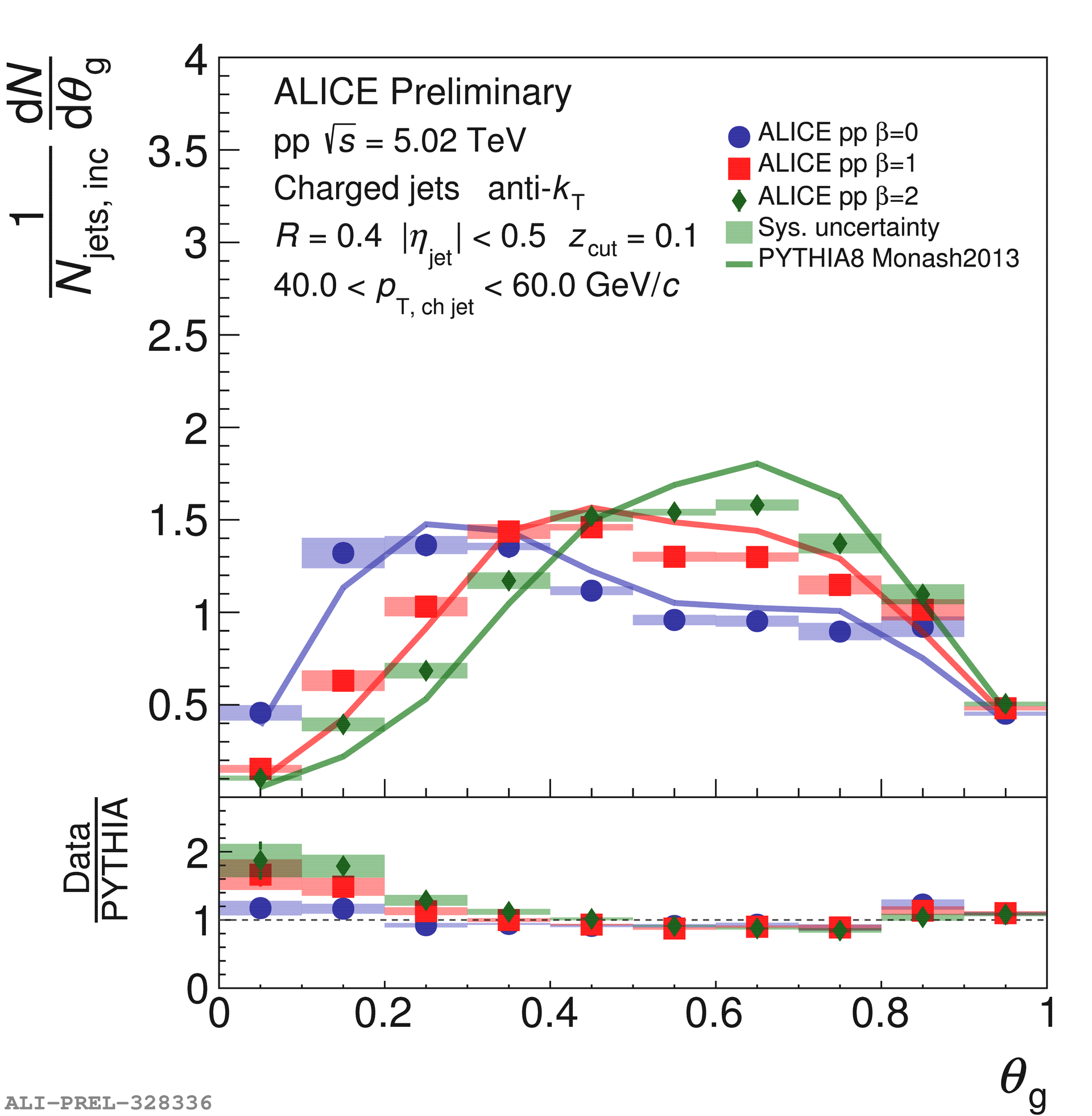}
	\caption{Charged-particle jet $z_g$ (left) and $\theta_g$ (right) in pp collisions at $\sqs=13$ TeV for different $\beta$ values, compared to PYTHIA 8 simulations.}
	\label{fig:chjet}
\end{figure}

Measurements of groomed jet substructures allow for access to the hard parton structure of a jet, while mitigating the effects of the underlying event and hadronization~\cite{Acharya:2019djg}. Ideally, it provides a direct interface with QCD calculations. 
Soft-drop grooming is a novel technique that is able to remove wide-angle soft radiation (such as initial-state radiation) as well as that of the underlying event~\cite{Larkoski:2014wba}.
In this method, the jets that had previously been reconstructed with the anti-\kT algorithm~\cite{Cacciari:2008gp} are first declustered and then reclustered using the Cambridge-Aachen algorithm~\cite{Dokshitzer:1997in} to form a clustering tree that follows angular ordering. Then the soft branches are iteratively removed if not fulfilling the so-called soft-drop condition, 
\vspace{-2mm}
\begin{eqnarray}
z > z_{\rm cut} \theta^\beta\,, & {\rm where} &  z = \frac{p_{{\rm T},2}}{p_{{\rm T},1}+p_{{\rm T},2}} \ {\rm and} \ \theta = \frac{\Delta R_{1,2} }{ R } 
\end{eqnarray}
are the momentum fraction taken by the second prong ($p_{{\rm T},1}$ and $p_{{\rm T},2}$ being the momenta of the two prongs), and the splitting radius (defined as the ratio of the $\Delta R_{1,2}$ splitting angle between the two prongs and the resolution parameter of the anti-\kT clustering). The soft threshold is set to $z_{\rm cut}=0.1$. The angular exponent $\beta$ is responsible for rejecting soft radiation.
The jet groomed momentum fraction $z_g$ and the groomed radius $\theta_g$, defined as the values of $z$ and $\theta$ corresponding to the first hard splitting fulfilling the soft-drop condition. 
Fig.~\ref{fig:chjet} shows $z_g$ (left panel) and $\theta_g$ (right panel) for charged-particle based jets for different choices of $\beta$.
For smaller $\beta$ values, more soft splittings are groomed away, leading to more collimated jets. This is clearly visible in the case of $\theta_g$ where the weight of the distribution shifts toward smaller angles with decreasing $\beta$. 

\begin{figure}
	\centering
	\begin{minipage}[c]{0.475\textwidth}
		\centering
		\includegraphics[width=0.8\textwidth]{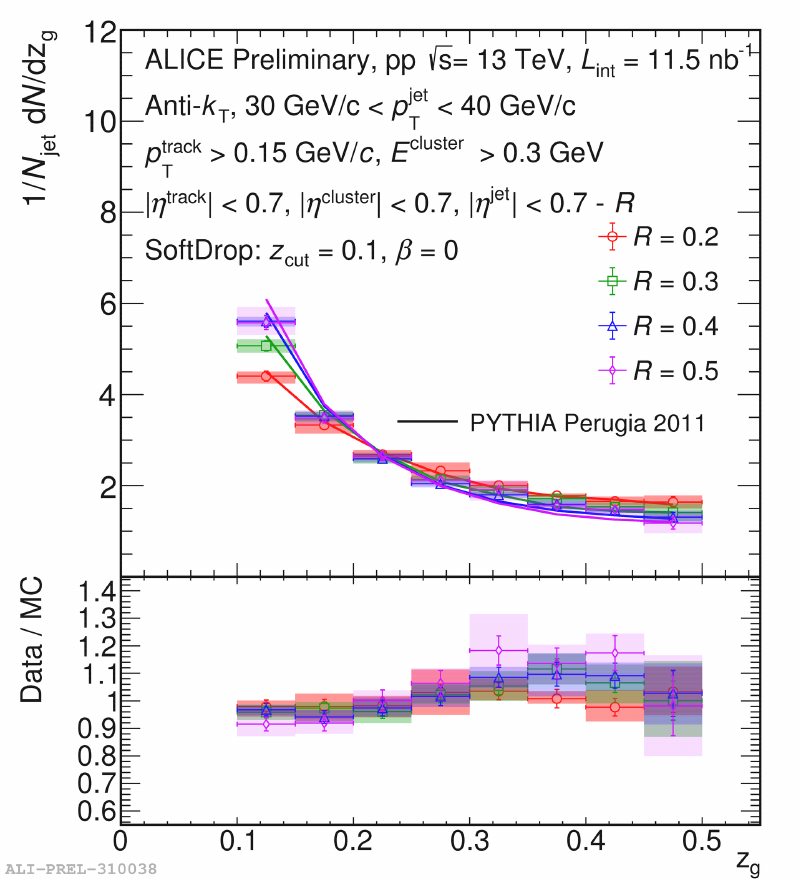}%
		\caption{\label{fig:fulljet_zg}Full jet $z_g$ for different jet resolution parameter values compared to PYTHIA simulations, in pp collisions at $\sqs=13$ TeV.}
	\end{minipage}
	\hspace\fill
	\begin{minipage}[c]{0.475\textwidth}
		\centering
		\includegraphics[width=.9\textwidth]{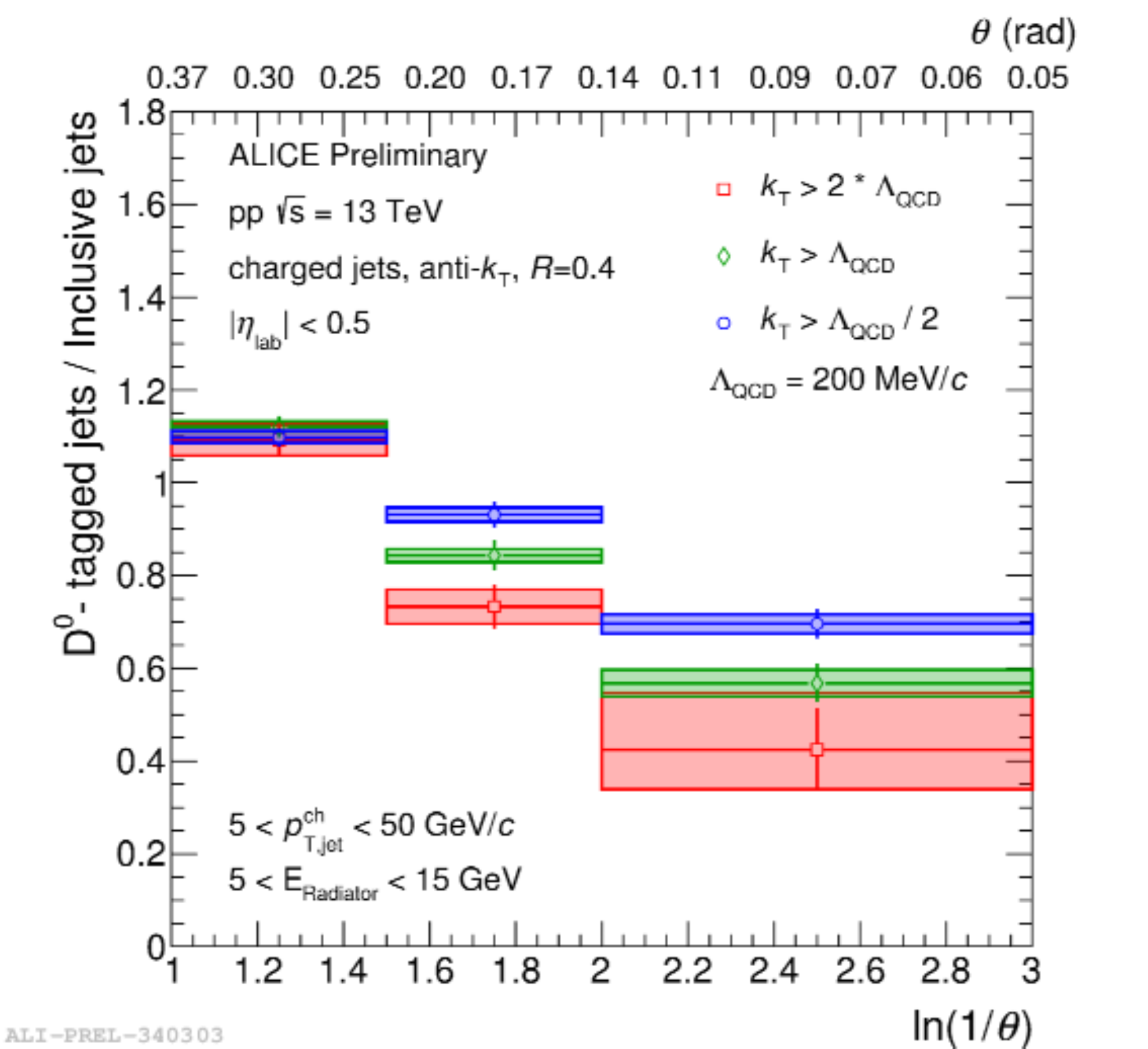}
		\caption{Ratio of the angular distribution of splittings with different $k_{\rm T}$ cuts for $D^0$-tagged jets over inclusive jets, shown for ${5<E_{\rm rad}<15}$~GeV.}
		\label{fig:deadcone}
	\end{minipage}
\end{figure}

Figure~\ref{fig:fulljet_zg} shows the full-jet groomed momentum fraction $z_g$ in the $30<p_{\rm T}^{\rm jet}<40$ GeV/$c$ transverse for different $R$ values. The difference suggests that jets with small radii tend to split more symmetrically, while in case of larger radii there is a higher sensitivity to non-perturbative effects. Trends observed both in the $R$ and the $\beta$-dependent groomed jet substructure results are reproduced rather well by Monte-Carlo event generators~\cite{Sjostrand:2007gs,Sjostrand:2006za}.

\section{Structure and fragmentation of heavy-flavor jets}

In gauge theories, charged particles with a mass $m>0$ and energy $E$ emit radiation that is suppressed below angles $\theta \approx m/E$ with respect to the axis of the radiator. This so-called dead-cone effect is expected to be present in jets containing heavy flavor~\cite{Dokshitzer:1991fd,Thomas:2004ie}. The ALICE collaboration presented the first direct measurement of the dead cone in heavy-flavor jets, following the iterative declustering method proposed in Ref.~\cite{Cunqueiro:2018jbh}. 
A cut on the relative transverse momentum fraction of the splitting, $k_{\rm T}$, is applied to remove non-perturbative effects.
Fig.~\ref{fig:deadcone} shows the ratio of the angular distribution of splittings for D$^0$-tagged jets over inclusive jets for radiator energy of ${5<E_{\rm rad}<15}$~GeV.
The D$^0$-tagged jets show a significant suppression toward smaller splitting angles. This suppression becomes stronger if the $k_{\rm T}$ cut is set to higher values, corresponding to a cleaner dead-cone signature with less contamination by non-perturbative effects.

\begin{figure}
	\centering
	\includegraphics[width=0.4\textwidth]{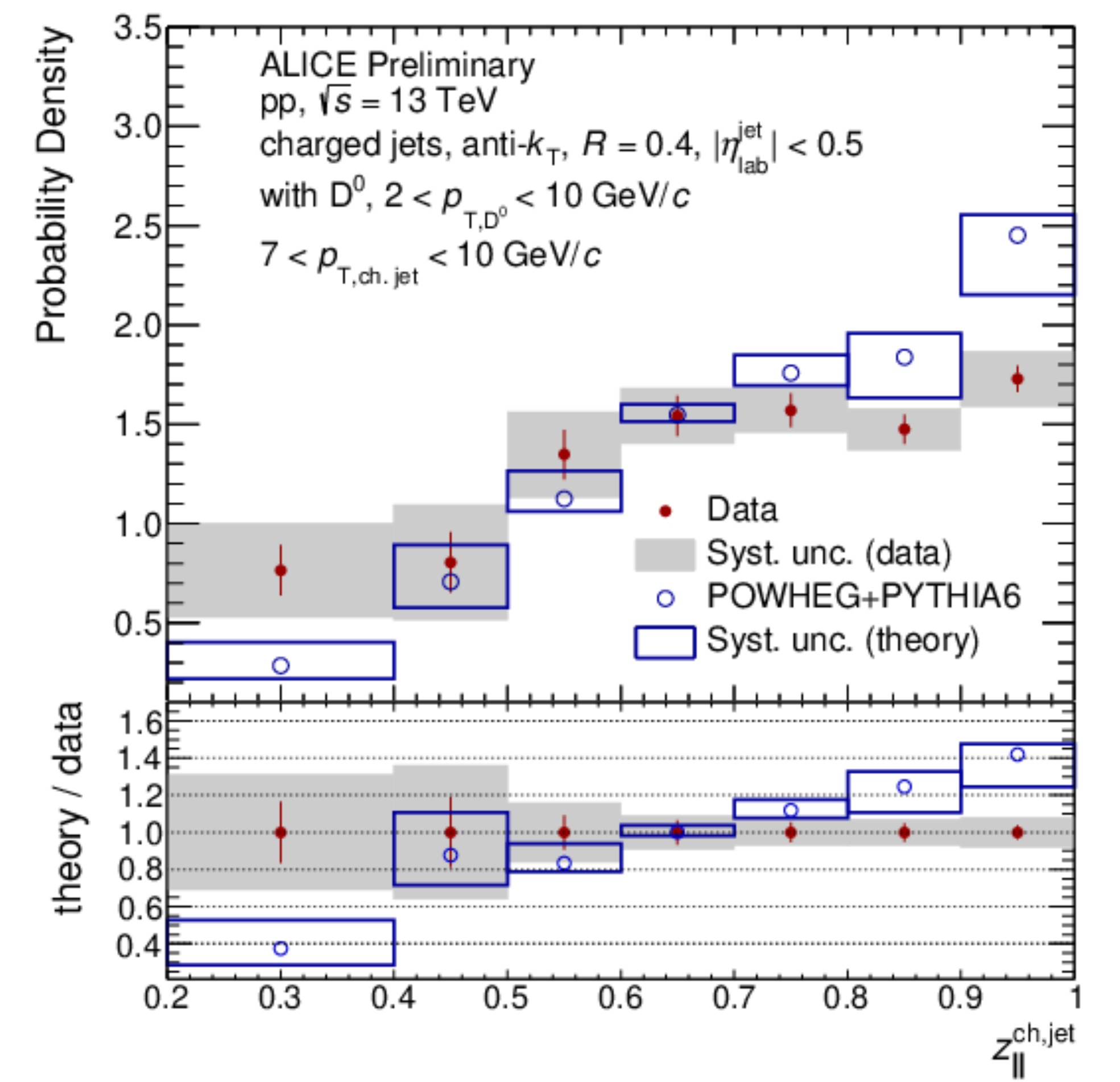}%
	\hspace{0.05\textwidth}
	\includegraphics[width=0.5\textwidth]{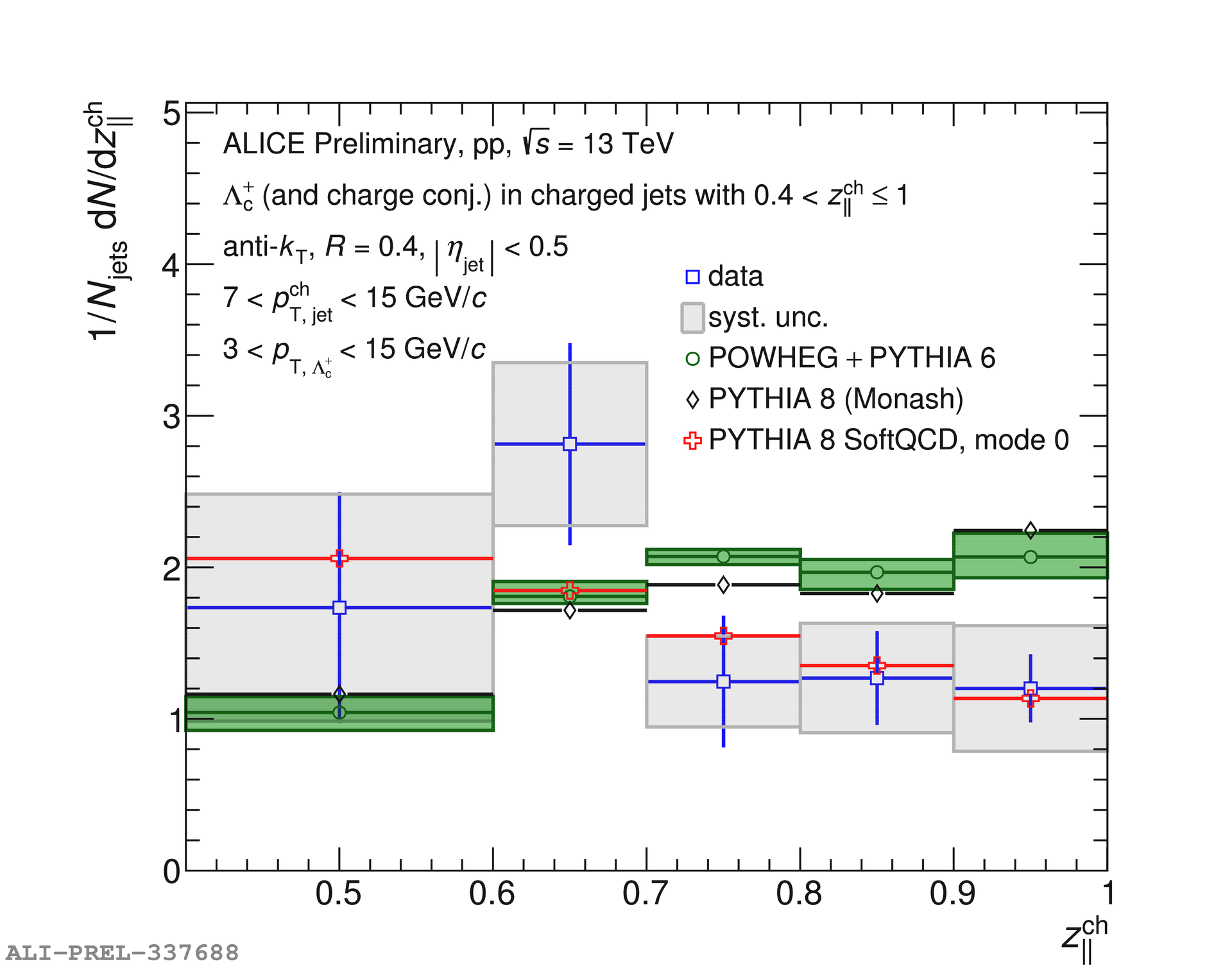}%
	\caption{Parallel momentum fraction $z_{\parallel}$ of charged-particle jets tagged with D$^0$ mesons (left) and \Lcp baryons (right).}
	\label{fig:zpar}
\end{figure}

The reconstruction of heavy-flavor hadrons within a jet allows for direct access to the fragmentation properties, and also allows for a comparison of meson and baryon fragmentation. The parallel momentum fraction $z_{\parallel}$ of D$^0$ mesons and \Lcp baryons, shown in Fig.~\ref{fig:zpar} left and right panels respectively, exhibit similar trends in the chosen momentum range. 
It is to be noted however, that a quantitative description of the observations still poses a challenge to some of the most popular model calculations~\cite{Sjostrand:2007gs,Alioli:2010xd,Christiansen:2015yqa}.

The groomed jet substructure of D$^0$-tagged jets has been measured for the first time, and compared to that of inclusive jets.
\begin{figure}
	\centering
	\includegraphics[width=0.33\textwidth]{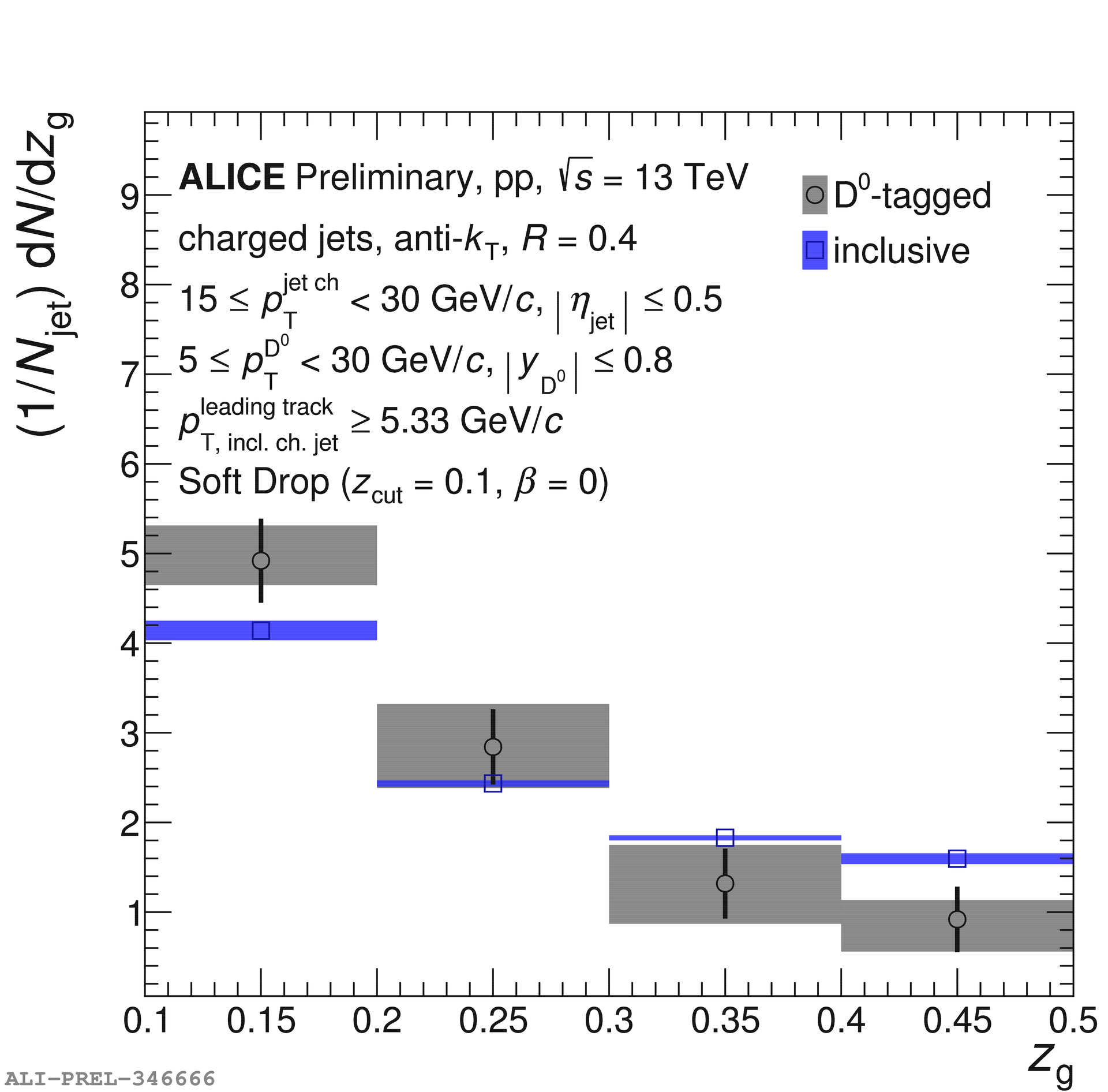}%
	\includegraphics[width=0.33\textwidth]{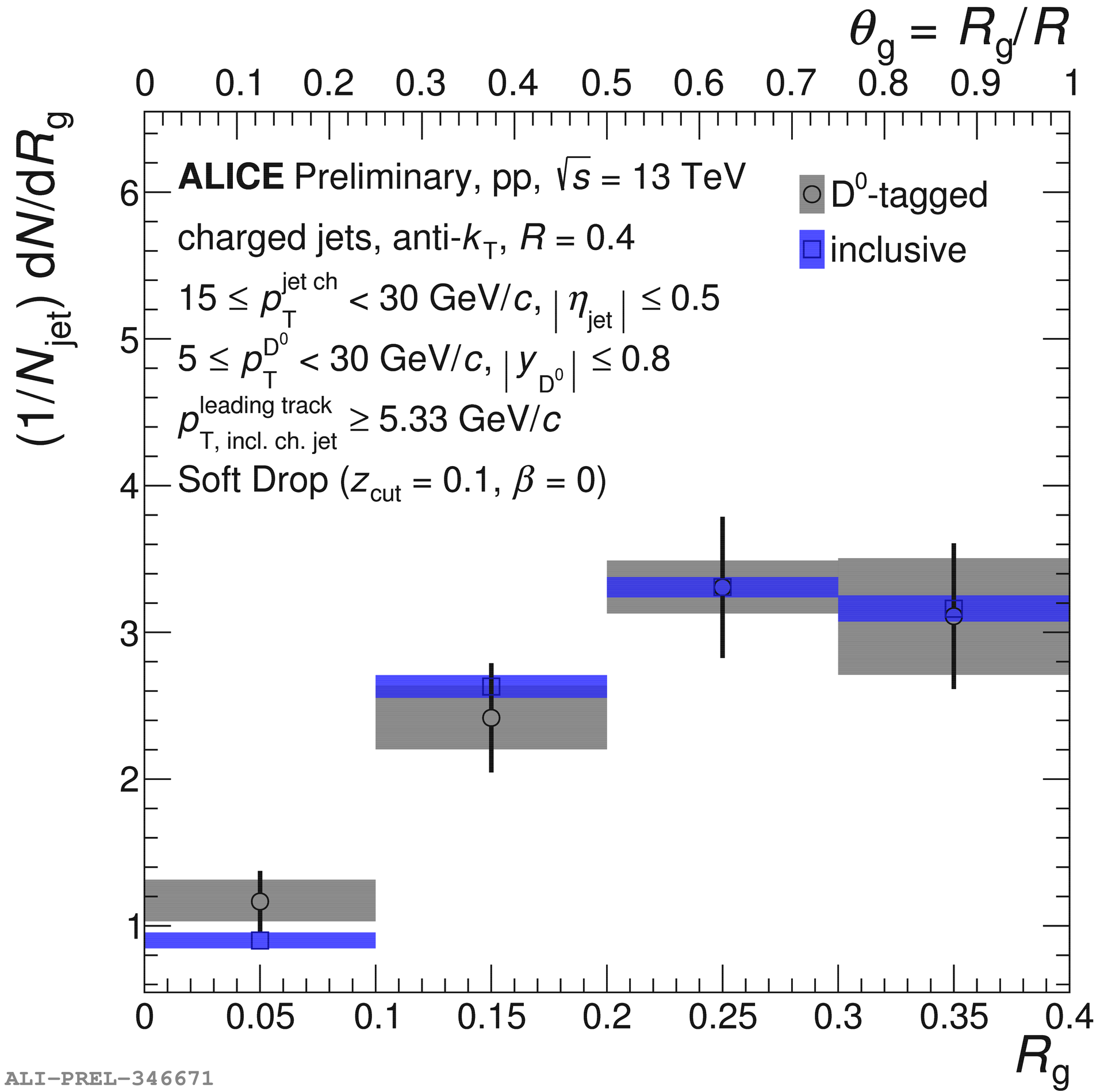}%
	\includegraphics[width=0.33\textwidth]{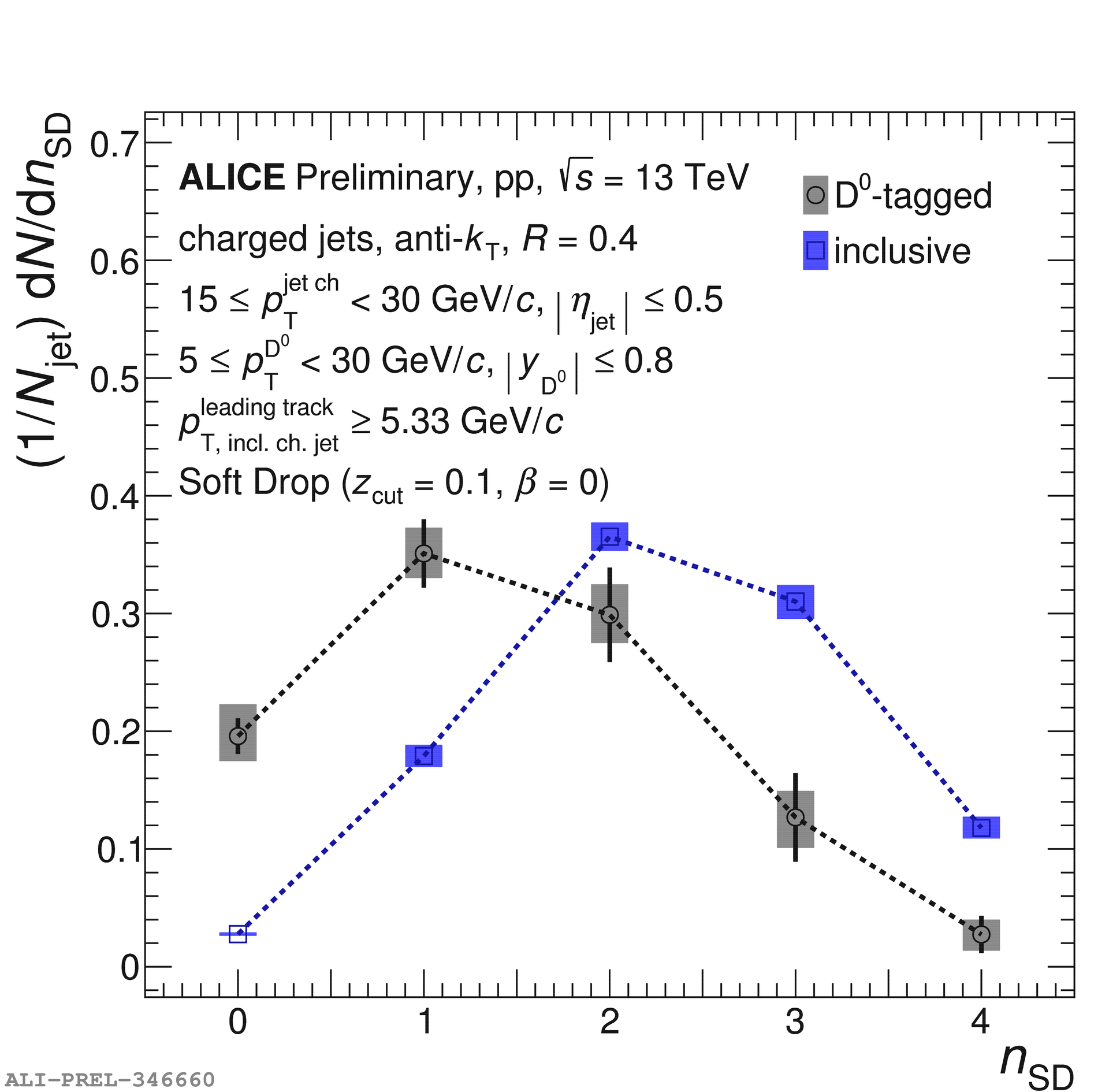}%
	\caption{\label{fig:dtagsubstr}Substructure variables $z_g$ (left), $\theta_g$ (center) and $n_{\rm SD}$ (right) of D$^0$-tagged charged-particle jets compared to inclusive charged-particle jets, in pp collisions at $\sqrt{s}=13$ TeV.}
\end{figure}
Trends in $z_g$ (Fig.~\ref{fig:dtagsubstr} left) and $\theta_g$ (Fig.~\ref{fig:dtagsubstr} center) are slightly different for charmed and inclusive jets, giving a hint about flavor-dependent jet substructure. A more obvious difference is present in the distribution of the number of splittings fulfilling the soft-drop condition, $n_{\rm SD}$ (Fig.~\ref{fig:dtagsubstr} right). The fact that charm jets typically have less hard splittings than inclusive jets is consistent with harder heavy-flavor fragmentation caused by mass and color charge effects.

\section{Summary}

In this contribution, recent jet-related results were presented from the ALICE experiment in pp collisions at $\sqrt{s}=13$ TeV. Soft-drop groomed substructure measurements of full and charged jets provide an excellent opportunity to test perturbative QCD and hadronization models, besides serving as a baseline for heavy-ion collisions. We also presented the first direct measurement of the dead cone in heavy-flavor jets. Parallel momentum fractions of charmed D$^0$ mesons and \Lcp baryons provide great discrimination power among models on heavy-flavor fragmentation. Charmed jets have been found to typically have less hard splittings than inclusive jets, suggesting a harder fragmentation of heavy than light flavor. The upcoming Run--3 phase of LHC with higher luminosity will allow for high-precision measurements of jets, charmed baryons as well as beauty-jets, further facilitating model developments and moving toward a deeper understanding of the strong interaction~\cite{Noferini:2018are}.

This work has been supported by the Hungarian NKFIH/OTKA K120660 and FK131979 grants, as well as the NKFIH 2019-2.1.6-NEMZ\_KI-2019-00011 project.

\end{document}